\documentclass[doublecol]{epl2} % for 2 columns style with line numbers
\usepackage{amssymb}

\usepackage[figuresright]{rotating}
\usepackage{graphicx}
\usepackage{mathrsfs}
\usepackage{bbm}
\usepackage{amssymb}
\usepackage{amsmath}
\usepackage{subfigure}
\usepackage{bm}
\usepackage{txfonts}
\newcommand{\ket}[1]{|#1\rangle}
\newcommand{\bra}[1]{\langle#1|}

\newcommand{\lr}\longrightarrow
\newcommand{\ra}\rightarrow

\title{Multipartite $d-$level GHZ bases associated with generalized braid matrices\footnote{G. Wang, C. Sun, C. Wu, B. Liu, Y. Zhang, and K. Xue, EPL (Europhysics Letters) 108, 10001 (2014).}}
\shorttitle{$d-$level multipartite GHZ bases} %Insert here a short version of the title if it exceeds 70 characters

\author{Gangcheng Wang\inst{1}\thanks{E-mail: \email{wanggc000@163.com}} \and Chunfang Sun\inst{1,2} \and Chunfeng Wu\inst{3} \and Bo Liu\inst{1} \and Yan Zhang\inst{1} \and Kang Xue\inst{1}\thanks{E-mail: \email{xuekang@nenu.edu.cn}}}
\shortauthor{Gangcheng Wang \etal}

\institute{
  \inst{1} School of Physics, Northeast Normal University, Changchun 130024, People's Republic of China\\
  \inst{2} Centre for Quantum Technologies, National University of Singapore, 3 Science Drive 2, Singapore 117543\\
  \inst{3} Pillar of Engineering Product Development, Singapore University of Technology and Design, 20 Dover Drive, 138682 Singapore
}

\abstract{
We investigate the generalized braid relation for an arbitrary multipartite $d$-level system and its application to quantum entanglement. By means of finite-dimensional representations of quantum plane algebra, a set of $d^{N}\times d^{N}$ unitary matrix representations satisfying the generalized braid relation can be constructed. Such generalized braid matrices can entangle $N-$partite $d-$level quantum states. Applying the generalized braid matrices on the standard basis of product states, one can obtain a set of maximally entangled bases. Further study shows that such entangled states can be viewed as the $N-$partite $d-$level Greenberger-Horne-Zeilinger (GHZ) states.}
\begin{document}
\maketitle
\section{Introduction}\label{secI}
One of the most prominent properties of quantum states is quantum entanglement, which has motivated much of work in quantum information theory \cite{nielsen2000}. For instance, quantum entangled channels are essential in teleporting an unknown quantum state \cite{bennett1993} and in sharing a secret key for cryptography \cite{bennett1992}. For qubits, multipartite entangled states have been studied intensively in quantum information processing including quantum nonlocality \cite{GHSZ}, one-way quantum computation \cite{raussendorf2001} and quantum error correcting codes \cite{calderbank1996,braunstein1998,cerf2002_1}, etc. Recently, investigations on these fields have been generalized to high-dimensional Hilbert spaces due to the enhanced security offered in quantum cryptography \cite{cerf2002_1}. One of the most important multipartite $d-$level (qudit) entangled states is the GHZ state \cite{cabello2001,cheong2007,cerf2002_2,lee2006}, since it is the maximally entangled multipartite state.

In mathematical physics, Yang-Baxter equation (YBE) is a fundamental tool to deal with quantum integrable models and statistical models \cite{yang1967,baxter1972}. As is well known, the parameter-dependent solution to YBE denoted by $\breve{R}(x)$ satisfies

\begin{equation}\label{YBE}
  \breve{R}_{12}(x)\breve{R}_{23}(xy)\breve{R}_{12}(y)=\breve{R}_{23}(y)\breve{R}_{12}(xy)\breve{R}_{23}(x),
\end{equation}

where $\breve{R}_{12}(x)\equiv \breve{R}(x)\otimes I$ and $\breve{R}_{23}(x)\equiv I\otimes \breve{R}(x)$ with $I$ being the identity matrix. The parameter-independent asymptotic form of $\breve{R}(x)$ denoted by $S$ obeys the braid relation

\begin{equation}\label{braid}
  S_{12}S_{23}S_{12}=S_{23}S_{12}S_{23},
\end{equation}

where $S_{12}\equiv S\otimes I$ and $S_{23}\equiv I\otimes S$. Recently, the unitary $\breve{R}(\theta)$ matrix as well as the unitary braid matrix $S$ have been investigated in quantum entanglement theory. Applying braid operator $S$ on bipartite product state $\ket{k,l}\equiv \ket{k}\otimes \ket{l}$, one gets a set of well-known maximal entangled states (\emph{i.e.}, the Bell basis) \cite{chen2007,xue2012,kauffman2004}. This provides a novel way to study quantum entanglement based on the theory of braiding operators, as well as YBE \cite{chakrabarti2013,ho2010,rowell2013,wang2006,wang2012,wang2010,rowell2012}. These studies include qubit case \cite{chen2007} and qutrit case \cite{wang2010,rowell2012}. In Refs. \cite{zhang2007,rowell2010}, the authors generalized the concept of braid group relation and YBE to multi-qubit systems, which are termed generalized YBE (gYBE). Very recently, Hastings \emph{et.al.} studied Gaussian representation based on the $p-group$ with odd prime $p$ \cite{jones1989} and applied the Gaussian representation to describe the braiding statistical behavior of Metaplectic anyons \cite{hastings2013,hastings2014}. Thus it is natural to ask whether the YBE approach can be generalized to an arbitrary multipartite $d$-level case, or the so-called $N-$qudit case, and whether one can get useful $N-$partite $d-$level entangled states via such approach.

Following Refs. \cite{zhang2007,rowell2010}, the aim of this paper is to generalize Yang-Baxter approach to multipartite $d$-level systems. We also explore the interesting applications of the generalization in the field of quantum entanglement. To achieve this, we first briefly review the theory of finite-dimensional representations of quantum plane algebra (QPA) \cite{zhou2003,ge1992} and then construct $N-$body $d-$level braid matrices by means of QPA, which provides the main mathematical tools for this work. We show that the $d-$level GHZ states for a $N-$body system can be obtained via unitary braid transformations, and the cases of $d=2$ and $d=3$ are discussed in detail as examples.

\section{Solution to the generalized braid matrix}\label{secII}
We begin with parameter-independent YBE (\emph{i.e.}, braid relation). Let $S^{(d)}_{1,\cdots,N}$ be the $N-$body $d-$level braid matrix; the following so-called generalized braid relation \cite{rowell2010} is satisfied

\begin{equation}\label{gbraid}
  S^{(d)}_{1,\cdots,N}S^{(d)}_{2,\cdots,N+1}S^{(d)}_{1,\cdots,N}=S^{(d)}_{2,\cdots,N+1}S^{(d)}_{1,\cdots,N}S^{(d)}_{2,\cdots,N+1}.
\end{equation}

When $N=2$, the generalized braid relation degenerates to its ordinary form (\emph{i.e.}, eq. (\ref{braid})). When $N=3$, the generalized $d-$level three-partite braid relation can be written as $S_{123}^{(d)}S^{(d)}_{234}S^{(d)}_{123}=S^{(d)}_{234}S^{(d)}_{123}S_{234}^{(d)}$.
Actually, the unitary generalized braid matrix $S^{(d)}_{1,\cdots,N}$ is a $N-$qudit quantum gate acting on the tensor product Hilbert space $\mathcal{V}_{1}^{d}\otimes \mathcal{V}_{2}^{d}\otimes \cdots \otimes \mathcal{V}^{d}_{N}$, where $\mathcal{V}_{i}^{d}$ is the $i^{th}$ vector space of dimension $d$. We show in the following that the multipartite $d-$level braid matrix can be contructed by resorting to QPA.

\subsection{Matrix representations of QPA}\label{secII:I}
In this subsection, we review some basic results about the finite-dimensional representations of QPA which provides us a useful mathematical tool to describe a qudit system. Then the solutions to the $N-$body $d-$level braid relation can therefore be constructed. The so-called QPA generated by $\{q, X, Z\}$, is defined by the following relation

\begin{equation}\label{Eq:QPA}
    XZ=q ZX,
\end{equation}

here $q$ is a complex number. It is well known that the associative algebra generated by $X$ and $Z$ possesses a $d-$dimensional irreducible representation only when $q^{d}=1$. In this paper, we take $q$ as a primitive $d^{th}$ root of unity $(i.e., q\equiv q_{d}\equiv e^{i2\pi/d})$. Such special case has been studied by Weyl \cite{weyl1932} and Schwinger \cite{schwinger1960}. Obviously, eq. (\ref{Eq:QPA}) implies $X^{m}Z^{n}=q^{mn} Z^{n}X^{m}$ for all integers $m$ and $n$.

If one takes $\{\ket{k}; k=0,1,\cdots, d-1\}$ as an orthogonal basis for one qudit Hilbert space $\mathcal{H}=\mathcal{V}^{d}$, the operators $X$ and $Z$ possess the corresponding realizations $X=\sum_{k=0}^{d-1}\ket{k\ominus 1}\bra{k}$ and $Z=\sum_{k=0}^{d-1}q^{k}\ket{k}\bra{k}$, and the matrix forms read

\begin{eqnarray}\label{Eq:XZ}
\begin{array}{c}
  X = \left(
          \begin{array}{ccccc}
            0 & 1 & 0 & \cdots & 0 \\
            0 & 0 & 1 & \cdots & 0 \\
            \vdots & \vdots & \ddots & \vdots & \vdots \\
            0 & 0 & 0 & \cdots & 1 \\
            1 & 0 & 0 & \cdots & 0 \\
          \end{array}
        \right),\\
         Z =\left(
           \begin{array}{ccccc}
             1 & 0 & \cdots & 0 & 0 \\
             0 & q & \cdots & 0 & 0 \\
             \vdots & \vdots & \ddots & \vdots & \vdots \\
             0 & 0 & \cdots & q^{d-2} & 0 \\
             0 & 0 & \cdots & 0 & q^{d-1} \\
           \end{array}
         \right).
\end{array}
\end{eqnarray}

Here and after we adopt a cyclic representation, which implies the relation $\ket{k\ominus l}\equiv \ket{(k-l)~mod~d}$. Acting $X^{l}$ and $Z^{l}$ on the state $\ket{k}$, one obtains $ X^{l}\ket{k}=\ket{k\ominus l}$, $Z^{l}\ket{k}=q^{kl}\ket{k}$. The results imply that the operators $X$ and $Z$ satisfy the relation $X^{d}=Z^{d}=I$ ($I$ is identity matrix). Obviously, when $d=2$, $q=-1$, the operators $X$ and $Z$ can be identified as Pauli matrices $\sigma_{x}$ and $\sigma_{z}$, respectively. From this viewpoint, the operators $X$ and $Z$ can be regarded as generalized Pauli matrices \cite{thas2009}. The operators $X$ and $Z$ are connected by $F$ with $X=F^{\dag}ZF$, where $F=d^{-1/2}\sum_{k,k'=0}^{d-1}q^{-kk'}\ket{k}\bra{k'}$ is the Fourier operator. Actually, the operators $X$, $Z$ and $F$ are often used in high-dimensional quantum error-correcting codes \cite{gottesman2001}.

\subsection{Matrix representations of $N-$partite $d-$level braid algebra}\label{secII:II}
In this subsection, we present the $N-$partite $d-$level braid matrix based on QPA. From QPA, we first obtain the generalized $M-$algebra. As is well known, $M-$algebra (or extra-special 2-group \cite{rowell2010}) obeying the relations $M^{2}=-I$ and $M_{12}M_{23}=-M_{23}M_{12}$, plays an important role in the theory of YBE \cite{chen2007}. By means of $M-$algebra, the braid matrix and $\breve{R}(x)$ can be constructed, and these matrix representations are applied to the studies of quantum entanglement and Berry phase. Now we show that the $M-$algebra can be generalized to the $N-$partite $d-$level system, and can be used to construct the generalized $M-$matrices (or extra-special $d$-group \cite{jones1989,hastings2013}).

To generalize the $M-$matrix, we introduce the matrices $A=ZX=\sum_{k=0}^{d-1}q^{k-1}\ket{k\ominus 1}\bra{k}$ and $B=X$ which satisfy the algebraic relations $A^{d}=(-1)^{d-1}I$, $B^{d}=I$ and $AB=q^{-1}BA$. Then the generalized $M-$matrix can be obtained in terms of matrices $A$ and $B$ with the following relation

\begin{equation}\label{gM}
  M^{(d)}=A\otimes B\otimes\cdots \otimes B,
\end{equation}

where $M^{(d)}$ is the $d-$level $M-$matrix, which obeys the following algebraic relations

\begin{eqnarray}\label{Eq:gMA}
\begin{array}{c}
 \left[M^{(d)}\right]^d =(-1)^{d-1}I,\\
 \\
 M^{(d)}_{1,\cdots,N}M^{(d)}_{2,\cdots,N+1} = qM^{(d)}_{2,\cdots,N+1}M^{(d)}_{1,\cdots,N},
\end{array}
\end{eqnarray}
here $M^{(d)}_{1,\cdots,N}$ stands for $M^{(d)}$ acting on the tensor product Hilbert space $\mathcal{H}^{N}_{d}=\mathcal{V}_{1}^{d}\otimes \mathcal{V}^{d}_{2}\otimes \cdots \otimes \mathcal{V}^{d}_{N}$.

With this generalized $M-$matrix, one can construct a corresponding generalized braid matrix, which is denoted by $S^{(d)}$. Its specific form reads

\begin{equation}\label{gS}
  S^{(d)}=\frac{1}{\sqrt{d}}\sum_{k=0}^{d-1}\omega^{k(k+1)}\left[M^{(d)}\right]^{k},
\end{equation}

with the parameter $\omega=e^{i\pi/d}$ with $d$ being even, and $\omega=e^{i2\pi/d}$ with $d$ being odd. The braid matrix $S^{(d)}$ is a $d^{N}\times d^{N}$ matrix acting on the tensor product space $\mathcal{H}^{N}_{d}$. Direct computation shows that the generalized braid matrix is unitary (\emph{i.e.}, $\left[S^{(d)}\right]^{\dag}=\left[S^{(d)}\right]^{-1}$). Actually, some special cases for $d$ and $N$ have been discussed in detail in Refs. \cite{chen2007,rowell2013,wang2010,rowell2012,jones1989}. When $d=N=2$, one obtain the so-called ``eight-vertex'' braid matrix, which can be viewed as localized representation of the Ising anyon (\emph{i.e.} the Majorana fermion) \cite{hastings2014} and can be used in the topological quantum computation theory \cite{hu2008}. For the case of odd prime $d\geq 3$ and $N=2$, substituting $M^{(d)}=\omega^{-1} M'^{(d)}$ into eq. (\ref{gS}), the braid matrix $S^{(d)}$ can be recast as $S^{(d)}=d^{-1/2}\sum_{k=0}^{d-1}\omega^{k^{2}}\left[M'^{(d)}\right]^{k}$. Such representation can be viewed as localization of the Gaussian representation, which is used in the Metaplectic anyons theory \cite{hastings2013,hastings2014,cobabera2014}. It is worth mentioning that the $N-$body braid matrices in eq. (\ref{gS}) are different from the braid matrices in Ref. \cite{yu2013,sun2012} since such $N-$body braid matrices cannot be factorized. That is to say the braid matrix in eq. (\ref{gS}) cannot be rewritten as direct product of local operators.
%In the following sections, we will utilize such approach to study $d-$level $N-$partite GHZ basis.

\section{Construction of $N-$partite $d-$level GHZ basis}\label{secIII}
In this section, we demonstrate that the generalized GHZ states can be constructed by resorting to the generalized braid matrix (\emph{i.e.}, eq. (\ref{gS})). Following Ge \emph{et al.}, one can view the generalized braid matrix as quantum gate acting on the tensor product Hilbert space $\mathcal{H}^{N}_{d}$. In other words, applying the generalized braid matrix on the standard basis, we are able to get a set of entangled states. For instance, if $d=N=2$, the Bell basis can be obtained. For the general case, one can achieve $N-$partite $d-$level entangled states.

The standard basis for the $N-$partite  $d-$level tensor product space reads $\mathcal{B}_{sta}=\{\ket{k_{1},k_{2},\cdots,k_{N}};k_{i}\in \{0,1,\cdots,d-1\}\}$. Acting $S^{(d)}$ on the standard basis state $\ket{k_{1},k_{2},\cdots,k_{N}}$, one easily gets

\begin{eqnarray}\label{N_gGHZ}
  &&\ket{\psi^{(d)}_{k_{1},k_{2},\cdots,k_{N}}} = S^{(d)}\ket{k_{1},k_{2},\cdots,k_{N}}\nonumber\\
  &&=\left\{\begin{array}{ll}
           \frac{1}{\sqrt{d}}\sum_{i=0}^{d-1}\alpha_{i}\ket{k_{1}\ominus i,k_{2}\ominus i,\cdots,k_{N}\ominus i}; & \mbox{odd $d$}\\
           \\
           \frac{1}{\sqrt{d}}\sum_{i=0}^{d-1}\beta_{i}\ket{k_{1}\ominus i,k_{2}\ominus i,\cdots,k_{N}\ominus i}; & \mbox{even $d$}
         \end{array}\right.
\end{eqnarray}

where $\alpha_{i}=\omega^{i(2k_{1}+i+1)/2}$ and $\beta_{i}=\omega^{2k_{1}i}$. Obviously, the state $\ket{\psi_{k_{1},k_{2},\cdots,k_{N}}^{(d)}}$ cannot be written as a product state $\ket{\psi_{1}^{(d)}}\otimes \ket{\psi_{2}^{(d)}}\otimes \cdots \otimes \ket{\psi_{N}^{(d)}}$, thus we can say the new basis $\mathcal{B}_{ent}=\{\ket{\psi^{(d)}_{k_{1},k_{2},\cdots,k_{N}}};k_{i}\in \{0,1,\cdots,d-1\}\}$ is an entangled basis. We next utilize an entanglement measure to test the degree of the entangled states. There are lots of entanglement measure for the multipartite system presented, such as genuine-multipartite-entanglement (GME) concurrence \cite{chen2012}, $n-$tangle \cite{wong2001} and so on. Here we adopt Scott's $Q-$measure \cite{scott2004}. Such entanglement measure was introduced to determine quantum entanglement for a multi-qudit state $\ket{\psi}$, which is a vector in the tensor product space $\mathcal{H}^{N}_{d}$. The so-called $Q-$measure reads

\begin{equation}\label{scott_measure}
  Q_{m}(\psi)\equiv \frac{d^{m}}{d^{m}-1}\left(1-\frac{m!(N-m)!}{N!}\sum_{|s|=m}\mbox{Tr}\rho_{s}^{2}\right),
\end{equation}

where $m=1,2,\cdots, \lfloor N/2 \rfloor$, $s\subset \{1,2,\cdots, N\}$ and $\rho_{s}=\mbox{Tr}_{s'}\ket{\psi}\bra{\psi}$ is the density operator for the qudits $s$ after tracing out the rest and the notation $\lfloor k\rfloor$ denotes the integer part of $k$. Scott showed that $Q_{m}$ is an entanglement monotone and $0\leq Q_{m}\leq1$. Substituting eq. (\ref{N_gGHZ}) into eq. (\ref{scott_measure}), we find that the basis states in eq. (\ref{N_gGHZ}) possesses the same entanglement degree. The general form of $Q_{m}(\ket{\psi^{(d)}_{k_{1},k_{2},\cdots,k_{N}}})$ is as follows

\begin{equation}\label{scott_measure_GHZ}
  Q_{m}(\ket{\psi^{(d)}_{k_{1},k_{2},\cdots,k_{N}}})\equiv 1-\frac{d^{m-1}-1}{d^{m}-1}.
\end{equation}

Specially, when $m=1$ the upper bound is reached (\emph{i.e.}, $Q_{1}(\ket{\psi^{(d)}_{k_{1},k_{2},\cdots,k_{N}}})\equiv 1$). According to Scott, the basis states in eq. (\ref{N_gGHZ}) are generalized GHZ states (\emph{i.e.}, $d-$level GHZ states). So we can conclude the new basis $\mathcal{B}_{ent}$ is a generalized $N-$partite GHZ basis. If we set $N=2$, one can get a set of $d-$level two-partite entangled states (\emph{i.e.}, the generalized Bell basis) \cite{klimov2009}.

To demonstrate our results explicitly, we consider the cases of $d=2$ and $d=3$ in the following subsections.

\subsection{The case of $N$ qubits}\label{IIIA}
When $d=2$, matrices $X$ and $Z$ are just Pauli matrices (\emph{i.e.}, $X=\sigma_{x}$ and $Z=\sigma_{z}$). The matrices $A$ and $B$ are as follows

\begin{equation}\label{d=2_AB}
  \begin{array}{cc}
    A=\left(
        \begin{array}{cc}
          0 & 1 \\
          -1 & 0 \\
        \end{array}
      \right),
     \qquad & B=\left(
        \begin{array}{cc}
          0 & 1 \\
          1 & 0 \\
        \end{array}
      \right)
  \end{array},
\end{equation}

here the matrices $A$ and $B$ satisfy the relation $AB=-BA$. By means of matrices $A$ and $B$, a $N-$partite two-level $M-$algebra can be constructed as $M^{(2)}=A\otimes B\otimes B \otimes\cdots \otimes B$, which satisfies the following algebraic relations

\begin{eqnarray*}
    [M^{(2)}]^{2}&=&-I,\\
  M^{(2)}_{1\ra N}M^{(2)}_{2\ra N+1}&=&- M^{(2)}_{2\ra N+1}M^{(2)}_{1\ra N}.
\end{eqnarray*}
From eq. (\ref{gS}), one can acquire the generalized braid group representation (BGR), $S^{(2)}=\frac{1}{\sqrt{2}}(I-M^{(2)})$, which fulfills the following $N-$body two-level braid relation

\begin{eqnarray*}
  S^{(2)}_{1\ra N}S^{(2)}_{2\ra N+1}S^{(2)}_{1\ra N} = S^{(2)}_{2\ra N+1}S^{(2)}_{1\ra N}S^{(2)}_{2\ra N+1}.
\end{eqnarray*}

If we let $N=2$, the specific form of the braid matrix $S^{(2)}$ is of the form

\begin{equation*}
  S^{(2)}=\frac{1}{\sqrt{2}}\left(
            \begin{array}{cccc}
              1 & 0 & 0 & -1 \\
              0 & 1 & -1 & 0 \\
              0 & 1 & 1 & 0 \\
              1 & 0 & 0 & 1 \\
            \end{array}
          \right).
\end{equation*}

The above braid matrix is the so-called ``eight-vertex'' braid matrix, which connects the standard basis $\{\ket{i,j};i,j=0,1\}$ with the Bell basis \cite{chen2007,kauffman2004}. Acting the $N-$partite two-level braid matrix $S^{(2)}$ on the standard basis, we can get a set of entangled states (\emph{i.e.} GHZ basis) having the same degree of entanglement. For example, acting $S^{(2)}$ on the product state $\ket{00\cdots 0}$, the standard GHZ state for the $N-$qubit system is thus found

\begin{eqnarray*}
  \ket{\psi_{00\cdots0}^{(2)}}=S^{(2)}\ket{00\cdots 0}=\frac{1}{\sqrt{2}}\left(\ket{00\cdots 0}+ \ket{11\cdots 1}\right).
\end{eqnarray*}

Our result for this special case is consistent with the ones obtained in Ref. \cite{zhang2007} by Ge \etal~ and in Ref. \cite{rowell2010} by Rowell \etal~.
\subsection{The case of $N$ qutrits}\label{IIIB}
When $d=3$, according to eq. (\ref{Eq:XZ}), the generators $X$ and $Z$ for $d=3$ can be obtained as follows

\begin{equation}\label{XZ3}
    \begin{array}{lr}
      X=\left(
          \begin{array}{ccc}
            0 & 1 & 0 \\
            0 & 0 & 1 \\
            1 & 0 & 0 \\
          \end{array}
        \right),
       & Z=\left(
             \begin{array}{ccc}
               1 & 0 & 0 \\
               0 & \omega & 0 \\
               0 & 0 & \omega^{2} \\
             \end{array}
           \right)
    \end{array}.
\end{equation}

By virtue of $X$ and $Z$, it is not difficult to find matrices $A$ and $B$, which satisfy the relations $A^{3}=B^{3}=I$ and $BA=\omega AB$.
The action of $A$ and $B$ on the standard basis of a qutrit system gives  $A\ket{i}=\omega^{i-1}\ket{i\ominus 1}$ and $B\ket{j}=\ket{j\ominus 1}$. The $N-$level $M-$matrix is therefore $M^{(3)}=A\otimes B\otimes B \otimes\cdots \otimes B$, which fulfills the following algebraic relations

\begin{eqnarray*}
\left[M^{(3)}\right]^{3}&=&I,\\
M^{(3)}_{1\ra N}M^{(3)}_{2\ra N+1}&=&\omega M^{(3)}_{2\ra N+1}M^{(3)}_{1\ra N}.
\end{eqnarray*}

Then one can acquire the three-level generalized BGR, $S^{(3)}=\frac{1}{\sqrt{3}}(I+\omega^{2} M^{(3)}+ \left[M^{(3)}\right]^{2})$ with the generalized braid relation satisfied, $S^{(3)}_{1\ra N}S^{(3)}_{2\ra N+1}S^{(3)}_{1\ra N} = S^{(3)}_{2\ra N+1}S^{(3)}_{1\ra N}S^{(3)}_{2\ra N+1}$.
We would like to mention that one special example of the $N$-qutrit case has been studied in detail in Refs. \cite{wang2010,rowell2012} with $N=2$. Acting $S^{(3)}$ on the standard basis of $N$ qutrits, it is easy to obtain a set of entangled states for tensor product space $\mathcal{H}_{3}^{N}=\mathcal{V}^{(3)}\otimes \mathcal{V}^{(3)}\otimes \cdots\otimes \mathcal{V}^{(3)} $. An example of this is to act the three-level $N-$body braid matrix  ($S^{(3)}$) on the product state $\ket{00\cdots0}$, and the following three-level GHZ-like state is found

\begin{equation}\label{Eq:GHZ-like}
 \ket{\psi_{00\cdots0}^{(3)}}=\frac{1}{\sqrt{3}}\left(\ket{00\cdots 0}+\ket{11\cdots 1} +\omega\ket{22\cdots 2}\right).
\end{equation}

And then by applying a local unitary operator $U=u\otimes u \otimes \cdots \otimes u$ with $u=\ket{0}\bra{0}+\ket{1}\bra{1}+\omega^{-1/N}\ket{2}\bra{2}$, one can verify that the phase factor $\omega$ in eq. (\ref{Eq:GHZ-like}) will vanish, and thus the standard $N-$partite $d-$level GHZ state can be obtained.

\section{Summary and discussion}\label{secIV}

In this paper, we have investigated the generalized braid relation ($N-$body $d-$level braid relation) and its applications to quantum entanglement. By means of finite-dimensional
matrix representations of QPA, a set of unitary matrix representations for the generalized braid relation can be constructed. We have showed that the generalized braid matrices are unitary, and such braid matrices can be viewed as quantum gates for $N-$body $d-$level systems. The action of the generalized braid matrix on the standard basis results in a set of entangled basis. The detailed calculations show that all the quantum states are $N-$partite $d-$level GHZ type states.

Let us make some discussions to end this paper. ($i$). In fact, we can always associate a Hamiltonian with the generalized unitary braid matrix. The unitary generalized braid matrix $S^{(d)}$ is of the form $S^{(d)}=\sum_{k}e^{i\varphi_{k}}\ket{u_{k}}\bra{u_{k}}$, where $e^{i\varphi_{k}}$'s and $\ket{u_{k}}$'s are the eigenvalues and eigenvectors of $S^{(d)}$. Then the corresponding Hamiltonian operator reads $H^{(d)}=-i$log$S^{(d)}$=$\sum_{k}\varphi_{k}\ket{u_{k}}\bra{u_{k}}$. This allows us to study braid transformation in a special physical system (such as NMR system). ($ii$). The generalized unitary braid matrices in this paper are spectral parameter-independent and time-independent. The braid matrices cannot be used to describe a parameter-dependent entangled basis and its dynamical properties. An exploration in the more complicated case with parameter dependence is still an open problem. As is known, Yang-Baxter equation can be viewed as spectral parameter-dependent braid relation. Via Yang-Baxterization approach, it is possible to generalize the parameter-independent braid matrix to the parameter-dependent $N-$partite $d-$level Yang-Baxter $\breve{R}$-matrix such that one can study the relation between $N-$partite $d-$level quantum entanglement and generalized YBE, and this is under investigation.

\acknowledgments
This work was supported by the NSF of China (Grant No. 11405026, No. 11175043, No. 11247005 and No. 11205028), the Plan for Scientific and Technological Development of Jilin Province (No. 20130522145JH) and the Jilin Post Doctorate Science Research Program (RB201330). CFS was also supported in part by the Government of China through CSC.


\begin{thebibliography}{0}

\bibitem{nielsen2000}
\Name{Nielsen M.A. \and Chuang I.L.}
\Book{Quantum Computation and Quantum Information}
\Publ{Cambridge University Press}
\Year{2000}.


\bibitem{bennett1993}
\Name{Bennett C.H., Brassard G., Crepeau C., Jozsa R., Peres A. \and Wootters W.K.}
  \REVIEW{Physical Review Letters}{70}{1993}{1895}.

\bibitem{bennett1992}
   \Name{Bennett C. H., Brassard G. \and Ekert A. K.}
  \REVIEW{Scientific American}{267}{1992}{50}.

\bibitem{GHSZ}
  \Name{Greenberger D.M., Horne M.A., Shimony A. \and Zeilinger A.}
  \REVIEW{Am. J. Phys.}{58}{1990}{1131}.

\bibitem{raussendorf2001}
  \Name{Raussendorf R. \and Briegel H.J. }
  \REVIEW{Physical Review Letters}{86}{2001}{5188}.

\bibitem{calderbank1996}
  \Name{Calderbank A.R. \and Shor P.W. }
  \REVIEW{Physical Review A}{54}{1996}{1098}.

\bibitem{braunstein1998}
  \Name{Braunstein S.L. }
  \REVIEW{Physical Review Letters}{80}{1998}{4084}.

\bibitem{cerf2002_1}
  \Name{Cerf N.J., Bourennane M., Karlsson A. \and Gisin N.}
  \REVIEW{Physical Review Letters}{88}{2002}{127902}.

\bibitem{cabello2001}
\Name{Cabello, A.}
\REVIEW{Physical Review A}{63}{2001}{022104}.

\bibitem{cheong2007}
  \Name{Cheong Y.W., Lee S.-W., Lee J. \and Lee H.-W. }
  \REVIEW{Physical Review A}{76}{2007}{042314}.

\bibitem{cerf2002_2}
  \Name{Cerf N.J., Massar S. \and Pironio S. }
  \REVIEW{Physical Review Letters}{89}{2002}{080402}.

\bibitem{lee2006}
  \Name{Lee J., Lee S.-W. \and Kim M.S. }
  \REVIEW{Physical Review A}{73}{2006}{032316}.

\bibitem{yang1967}
   \Name{Yang C.N.}
  \REVIEW{Physical Review Letters}{19}{1967}{1312}.

\bibitem{baxter1972}
  \Name{Baxter R.J.}
  \REVIEW{Annals of Physics}{70}{1972}{193}.

\bibitem{chen2007}
  \Name{Chen J.-L., Xue K. \and Ge M.-L. }
  \REVIEW{Physical Review A}{76}{2007}{042324}.

\bibitem{xue2012}
  \Name{Xue K. \and GE M.-L.}
  \REVIEW{International Journal of Modern Physics B}{26}{2012}{1243007}.

\bibitem{kauffman2004}
  \Name{Kauffman L.H. \and Lomonaco S.J. }
  \REVIEW{New Journal of Physics}{6}{2004}{134}.

\bibitem{chakrabarti2013}
  \Name{Chakrabarti A., Chakraborti A. \and Hidalgo E.G.}
  \REVIEW{Journal of Mathematical Physics}{54}{2013}{013517}.

\bibitem{ho2010}
  \Name{Ho C.-L., Solomon A.I. \and Oh C.-H.}
  \REVIEW{EPL (Europhysics Letters)}{92}{2010}{30002}.

\bibitem{rowell2013}
  \Name{Galindo C. \and Rowell E. C.}
  \REVIEW{J.Math.Phys.}{55}{2014}{061702}.

\bibitem{wang2006}
  \Name{WANG Z., FRANKO J.M. \and ROWELL E.C. }
  \REVIEW{Journal of Knot Theory and Its Ramifications}{15}{2006}{413}.

\bibitem{wang2012}
  \Name{Wang G., Xue K., Sun C. \and Du G. }
  \REVIEW{Quantum Inf Process}{11}{2012}{1775}.

\bibitem{wang2010}
  \Name{Wang G., Xue K., Sun C., Zhou C., Hu T. \and Wang Q. }
  \REVIEW{Quantum Inf Process}{9}{2009}{699}.

\bibitem{rowell2012}
  \Name{Rowell E. \and Wang Z. }
  \REVIEW{Commun. Math. Phys.}{311}{2012}{595}.

\bibitem{zhang2007}
\Name{Zhang, Y. \and Ge M.-L.}
\REVIEW{Quantum Information Processing}{6}{2007}{363}.

\bibitem{rowell2010}
  \Name{Rowell E. C., Zhang Y., Wu Y. S. \and Ge M. L. }
  \REVIEW{Quantum Inf. Comput.}{10}{2010}{685}.

\bibitem{jones1989}
  \Name{Jones V.F.R.}
  \REVIEW{Commun. Math. Phys.}{125}{1989}{459}.


\bibitem{hastings2013}
  \Name{Hastings M., Nayak C. \and Wang Z. }
  \REVIEW{Physical Review B}{87}{2013}{165421}.

\bibitem{hastings2014}
  \Name{Hastings Matthew B., Nayak Chetan \and Wang Zhenghan}
  \REVIEW{Communications in Mathematical Physics}{330}{2014}{45}.


\bibitem{zhou2003}
  \Name{Zhou D., Zeng B., Xu Z. \and Sun C.}
  \REVIEW{Physical Review A}{68}{2003}{062303}.

\bibitem{ge1992}
  \Name{Ge Mo-Lin, Liu Xu-Feng \and Sun Chang-Pu}
  \REVIEW{Journal of Physics A: Mathematical and General}{25}{1992}{2907}.

\bibitem{weyl1932}
  \Name{Weyl H.}
  \Book{Theory of Groups and Quantum Mechanics}
  \Publ{E. P. Dutton Co., New York}
  \Year{1932}

\bibitem{schwinger1960}
  \Name{Schwinger J.}
  \REVIEW{Proc. Natl. Acad. Sci. U.S.A.}{46}{1960}{570}.

\bibitem{thas2009}
  \Name{Thas K.}
  \REVIEW{EPL (Europhysics Letters)}{86}{2009}{60005}.

\bibitem{gottesman2001}
  \Name{Gottesman D., Kitaev A. \and Preskill J.}
  \REVIEW{Physical Review A}{64}{2001}{012310}.

\bibitem{hu2008}
  \Name{Hu S.-W., Xue K. \and Ge M.-L.}
  \REVIEW{Physical Review A}{78}{2008}{022319}.

\bibitem{cobabera2014}
  \Name{Cobanera E. \and Ortiz G.}
  \REVIEW{Physical Review A}{89}{2014}{012328}.

\bibitem{yu2013}
  \Name{Yu L.-W., Zhao Q. \and Ge M.-L.}
  \REVIEW{Annals of Physics(N. Y.)}{348}{2013}{106}.

\bibitem{sun2012}
  \Name{Sun C., Xue K., Wang G., Zhou C. \and Du G. }
  \REVIEW{Quantum Inf Process}{11}{2012}{385}.

\bibitem{chen2012}
  \Name{Chen Z.-H., Ma Z.-H., Chen J.-L. \and Severini S.}
  \REVIEW{Physical Review A}{85}{2012}{062320}.

\bibitem{wong2001}
  \Name{Wong A. \and Christensen N.}
  \REVIEW{Physical Review A}{63}{2001}{044301}.

\bibitem{scott2004}
  \Name{Scott A. J.}
  \REVIEW{Physical Review A}{69}{2004}{052330}.

\bibitem{klimov2009}
  \Name{Klimov A. , Sych D., Sanchez-Soto L. \and Leuchs G.}
  \REVIEW{Physical Review A}{79}{2009}{052101}.

\end{thebibliography}
\end{document}